\documentclass[aps, prab, reprint, showpacs, amsmath, amssymb]{revtex4-1}

\usepackage{amssymb}
\usepackage{amsmath}
\usepackage{amscd}
\usepackage{bm}
\usepackage{color,soul}
\usepackage[english]{babel}
\usepackage{graphicx}
\usepackage{hyperref}
\usepackage{booktabs}
\usepackage{color,soul}
\usepackage{dcolumn}
\usepackage{physics}
\usepackage{wrapfig}

\begin{document}
\title{ Self amplification of channeling radiation \\ 
at Bragg diffraction conditions }

\author{Stasis Chuchurka$^1$}%
\email{s.chuchurka@outlook.com}

\author{Andrei Benediktovitch$^{1,2}$}%
\email{andrei.benediktovitch@cfel.de}

\author{Aliaksandr Leonau$^{1}$}%

\author{Sergey Galyamin$^{3}$}%

\affiliation{$^1$Belarusian State University, 4 Nezavisimosti Ave., Minsk, 220030 Belarus}%
\affiliation{$^2$CFEL, DESY, 85 Notkestrasse, Hamburg, 22607 Germany}
\affiliation{$^3$Saint Petersburg State University, 7/9 Universitetskaya nab., St. Petersburg, 199034 Russia}

\date{\today}
\begin{abstract}
X-ray radiation of relativistic electrons passing through the crystal lattice (in case of channeling radiation) requires electrons with energies just from tens up to hundreds MeV, but has a drawback: low number of emitted X-ray quanta. However, the brightness of such a source could be potentially increased if the current density of the electron bunch is high enough to initiate the self amplified spontaneous emission (SASE) process. The conditions at which this phenomenon could take place for the case of axial channeling are analyzed in the present paper within the first order perturbation theory. The transition from spontaneous to SASE regime is described, the requirements for bunch current parameters initiating SASE process are determined taking into account the periodic structure of the crystal medium. It is also shown that satisfying Bragg diffraction conditions could enhance self amplification. Numerical results for the case of axial channeling in Si, Ge and C crystals are presented as an example. 

\end{abstract}

\maketitle


\section{\label{sec:introduction}Introduction}
The construction of the 4th-generation X-ray sources -- the X-ray Free Electron Lasers (XFEL) -- opened the unique possibilities to investigate matter on the time scale of femtoseconds with atomic spacial resolution.
As an example of such new revolutionary discoveries using XFELs one can mention  determination of the protein nanostructures \cite{Fromme2015} and description of fs-dynamics of chemical reactions \cite{Wernet2015,Minitti2015}. However, due to the significant cost of construction as well as the complexity of operation and maintenance (the size of XFELs can reach several km) currently only few installations of this type are operating (LCLS in the USA, SACLA in Japan, PAL-FEL in South Korea, eXFEL in Germany and Swiss-FEL in Switzerland), which makes the unique properties of XFELs inaccessible to the wider scientific community. In the view of this it is extremely important and relevant today to consider the problem of creating a compact source of bright and short X-ray pulses \cite{Corde2013}.  

This problem can be divided into the two objectives: (i) the search for a compact accelerator allowing to obtain ultrashort ($\sim$ 10--100 fs) electron bunches with low emittance; (ii) creation of a compact radiator.
Nowadays there exist several suggestions on using such versions of compact accelerators as: linear THz accelerators \cite{Nanni2015}, dielectric laser accelerators \cite{England2014}, laser wakefield accelerators \cite{Vieira2014,Ju2014,Li2014} and plasma wakefield accelerators \cite{Joshi2010,Lotov2014,Viera2014plasma,Blumenfeld2007}. As a compact radiator one can consider using the cryogenic undulators \cite{Oshea2010}, ``crystal'' undulators \cite{channeling2013} and others. Besides undulator radiation mechanism, there are other ways of generating  the X-ray radiation that at some specific conditions can also lead to SASE process. Among them: parametric X-ray radiation (PXR) \cite{Barysh2005}, channeling radiation (CR) \cite{Brau2012}, Cherenkov radiation near K-edge \cite{Knulst2004} and others \cite{elOfElectrons}. It is important to stress that electron energies required by these schemes can be reduced to tens of MeV whereas for XFELs this value reaches $\sim$ 3-20 GeV. The principal drawback of the sources based on these radiation mechanisms is low number of emitted X-ray quanta which is conditioned by limitations on interaction lengths and current densities. However, the brightness of the source increases drastically in cases when electrons radiate coherently.

The coherence in radiation between electrons can be achieved in two ways: (i) the electron bunch can be modulated in advance with modulation period equal to the radiation wavelength \cite{bunched}, or (ii) the electron beam can become bunched itself due to the interaction with the generated radiation by it \cite{ferbar}.

In the present paper we consider the possibility of initiating the SASE process starting from the spontaneous emission in the case of CR. The first investigation of this phenomenon was considered in \cite{ferbar} and later on developed in \cite{bardub1,bardub2,strauss1,strauss2}. However, the main focus in the aforementioned papers is set on the ways of reducing the current density threshold values in order to start the laser generation. We extend the ideas mentioned in these papers but take into account that technical progress in the field of generating ultrashort electron bunches makes the needed properties (high current density, low emittance  and big interaction length)   feasible \cite{beams}. Also, the ``diffraction before destruction'' concept \cite{destruction} which in our case can be reflected as ``radiation before destruction'' concept removes restrictions on  current density in the case of fs electron bunches. Description of the initial stage of SASE process in case of CR and analysis of the conditions at which SASE can be observed is done within the linear response theory. In order to favor the condtitions for SASE onset and reduce the requirements for the beam we benefit from the crystallographic order in a way that it forms the distributed feedback  which is known to improve the lasing properties \cite{feedback}. In this paper we consider the cubic crystal family and [001] axial channeling.

The paper is organized as follows. In section \ref{sec:firstpr} we perform the first principle derivation of the Hamiltonian density of the considered model. In section \ref{sec:2level} we reduce the CR bound state problem to the two-level system approximation. In section \ref{sec:temp} we describe the initial stage of SASE process within the perturbation theory that allows us to analyze the dispersion equation in section \ref{sec:disp} and boundary conditions in section \ref{sec:boundary}. In section \ref{sec:optimization} we perform optimization of conditions at which the considered effect could be observed. Finally, we present numerical results for the case of axial channeling in Si, Ge and C crystals in section \ref{sec:num}.

\section{\label{sec:firstpr}First principle formalism}
\subsection{\label{sec:quantization}Field quantization}

In order to describe the mechanism of SASE, quantization of the electromagnetic field inside the media should be performed. 
In this paper we use $\bm{\nabla}\cdot \left(\varepsilon\textbf{A}\right) = 0$ gauge. Consider the following Lagrangian density:

\begin{equation}
    \label{eqn:fieldLag}
    \mathcal{L}_f=\frac{1}{8\pi c^2}(\mathbf{\dot A \varepsilon \dot A} - c^2 (\bm{\nabla\times \mathbf{A}})^2),
\end{equation}

\noindent where $\mathbf{A}$ is the electromagnetic field vector-potential, $\varepsilon$ is the permittivity of the media. Expression (\ref{eqn:fieldLag}) can be used to obtain the generalized momentum

\begin{equation}
    \label{eqn:P}
    \bm{\mathcal{P}}=\frac{\partial\mathcal{L}_f}{\partial\mathbf{\dot A}}=\frac{\mathbf{\varepsilon \dot A}}{4\pi c^2}
\end{equation}

\noindent and Hamiltonian density:

\begin{equation}
    \label{eqn:fieldHam}
    \mathcal{H}_f=2\pi c^2\bm{\mathcal{P}}  \varepsilon^{-1} \bm{\mathcal{P}} + \frac{1}{8\pi} (\bm{\nabla\times \mathbf{A}})^2.
\end{equation}

Hamilton's equations derived from (\ref{eqn:fieldHam}) generate the wave equation for $\mathbf{A}$:

\begin{equation}
\begin{split}
    \label{eqn:HamEqns}
    \bm{\dot\mathcal{P}}=-\frac{\partial{H_f}}{\partial \mathbf{A}}=-\frac{1}{4\pi}\bm{\nabla}\cross \left( \bm{\nabla}\cross\mathbf{A} \right),\\
     \mathbf{\dot A}=\frac{\partial{H_f}}{\partial \bm{\mathcal{P}}}=4\pi c^2 \varepsilon^{-1}\bm{\mathcal{P}}.
\end{split}
\end{equation}

In order to quantize the electromagnetic field one should replace both vectors $\mathbf{A}$ and $\bm{\mathcal{P}}$ with the corresponding quantum operators and choose the following canonical commutation relations

\begin{equation}
\begin{split}
    \label{eqn:ComRels}
    &\left[ A_i(\mathbf{r},t),\mathcal{P}_j(\mathbf{r'},t)  \right] = i \hbar \delta_{ij}^{(tr)} (\mathbf{r-r'}),\\
    &\left[ \textbf{A}(\mathbf{r},t),\textbf{A}(\mathbf{r'},t)  \right] = 0,\\
    &\left[ \bm{\mathcal{P}}(\mathbf{r},t),\bm{\mathcal{P}}(\mathbf{r'},t)  \right] = 0.
\end{split}
\end{equation}

It is important to stress that vectors $\mathbf{A}$ and $\bm{\mathcal{P}}$ are transverse due to the gauge we utilize in the present paper.

For the further investigation it is convenient to perform transition to the Fourier ($\mathbf{k}, \omega$) space:

\begin{equation}
\begin{split}
    \label{eqn:fouTra}
    &\mathbf{A}(\mathbf{r},t)=\int\mathbf{A}(\mathbf{k},\omega)\exp \left[ i(\mathbf{k}\mathbf{r}-\omega t) \right]d\mathbf{k}d\omega.
\end{split}
\end{equation}

\subsection{\label{sec:channeling}Axial channeling}

In order to introduce notations we provide the reader with a brief description of the axial channeling \cite{bird}. Let us consider a relativistic electron moving close to an arbitrary crystallographic axis. In such a case the motion of an electron can be bounded by the electric potential of atoms of the crystal so that channeling phenomenon takes place. The corresponding Dirac equation describing the moving electron which interacts with atoms of the crystal has the following form:

\begin{align}
    \label{eqn:eleHam}
    H_e=m c^2\beta+c\bm{\alpha}\bm{p}- e V(\textbf{r}), \\
    \beta= 
    \begin{pmatrix}
    1 & 0 \\
    0 & -1
    \end{pmatrix}
    , \quad \bm{\alpha}= 
    \begin{pmatrix}
    0 & \bm{\sigma} \\
    \bm{\sigma} & 0
    \end{pmatrix}
    , \nonumber
\end{align}

\noindent where $\bm{\sigma}$ is a vector composed of Pauli matrices, $V$ is an electric potential produced by atoms of the crystal. In our case $V$ can be defined as an average potential of  atoms forming the crystallographic axis.

$H_e$ can be diagonalized by solving the eigen problem:

\begin{equation}
    \label{eqn:eigEqn}
    \begin{pmatrix}
    m c^2-eV-E_e & c\bm{\sigma} \mathbf{p} \\
    c\bm{\sigma} \mathbf{p} & -m c^2-eV-E_e
    \end{pmatrix}
    \begin{pmatrix}
    \phi \\
    \psi
    \end{pmatrix}
    =0.
\end{equation}

The system of equations (\ref{eqn:eigEqn}) allows one to establish relation between $\phi$ and $\psi$ functions:

\begin{equation}
    \label{eqn:psiEqn}
    \psi=\frac{c\bm{\sigma}\mathbf{p}\phi}{mc^2+E_e+eV(\textbf{r})},
\end{equation}

\noindent so that equation involving only one of them (for instance, $\phi$) can be derived:

\begin{equation}
    \label{eqn:phiEqn}
    (m c^2-eV(\textbf{r})-E_e)\phi+c\bm{\sigma}\mathbf{p}\frac{c\bm{\sigma}\mathbf{p}\phi}{mc^2+E_e+eV(\textbf{r})}=0.
\end{equation}

Let us define $p_\parallel$ as the projection of $\mathbf{p}$ on the direction of crystallographic axis and $\mathbf{p_\bot}$ as the remaining part of $\mathbf{p}$. One should note that $H_e$ commutes with $p_\parallel$ and, hence, both of them can be defined simultaneously. The atomic potential is much smaller than the rest mass Taking into account that atomic potential is much smaller than the rest mass of the electron where as its total energy in the relativistic case is much greater it (i.e., $eV \ll mc^2 \ll E_e$), expressions (\ref{eqn:psiEqn}) and (\ref{eqn:phiEqn}) can be reduced to (the Schr\"{o}dinger picture is considered below):

\begin{equation}
\begin{split}
    \label{eqn:chaEqns}
    &\frac{\hbar^2\Delta_\bot}{2\gamma m}\phi+(eV(\textbf{r})+\delta E)\phi=0,\\
    &\psi=\frac{\bm{\sigma}\mathbf{p}\phi}{\gamma mc},
\end{split}
\end{equation}

\noindent where $\delta E=E_e-c\sqrt{m^2c^2+p^2_\parallel}$ and $\gamma$ is associated with the momentum $p_\parallel$. It is remarkable that the transversal motion can be described by the equation which is similar to non-relativistic Schr\"{o}dinger equation (the only difference consists in multiplying the mass by $\gamma$). Solving  it one can obtain the critical angle at which electrons become channeled:

\begin{equation}
\begin{gathered}
    \label{eqn:critical}
    \theta_c\approx\sqrt{\frac{2V_0}{\gamma m c^2}},
\end{gathered}
\end{equation}

\noindent where $V_0$ is the depth of the potential well. Finally, the electron's state can be described by eigenvectors from (\ref{eqn:chaEqns}) and $H_e$ turns into

\begin{equation}
    \label{eqn:eleHamD}
    H_e =\sum_{n,l} \int E_{p,n}  \ket{p,n,l}\bra{p,n,l}dp,
\end{equation}

\noindent where $\ket{p,n,l}$ is a ket vector describing the electron's state with the definite projection of momentum $p_\parallel=p$ and energy state number $n$; $l$ is the index accounting the states with the same energy. In the coordinate representation $\ket{p,n,l}$ can be written in the following form:

\begin{equation}
    \label{eqn:eKet}
    \ket{p,n,l}=\frac{1}{\sqrt{2}} \exp{\frac{i H_e t}{\hbar}} 
    \begin{pmatrix}
        1\\
        \frac{\bm{\sigma}\mathbf{p}}{\gamma mc}
    \end{pmatrix}
    \exp{\frac{ipr_\parallel}{\hbar}}\phi_{n,l}(\textbf{r}_\bot).
\end{equation}

\subsection{\label{sec:interaction}Interaction}

The part of the Hamiltonian density describing the interaction between the electrons and the electromagnetic field (neglecting the Coulomb interaction between electrons) can be written in the following form:

\begin{equation}
    \label{eqn:intHam}
    \mathcal{H}_{int}=e\sum_j\bm{\alpha}_j\delta(\mathbf{r}_j-\textbf{r})\textbf{A}(\textbf{r}), 
\end{equation}

\noindent where summation is performed over the electron in the bunch. The presence of the interaction part in Hamiltonian density adds a new term to the wave equation:

\begin{equation}
    \label{eqn:current}
    \frac{i}{\hbar}\left[\mathcal{H}_{int},\bm{\mathcal{P}}\right]=-e\sum_j\bm{\alpha}_j\delta(\textbf{r}_j-\textbf{r}). 
\end{equation}

As a result, let us introduce the current density into the full wave equation:

\begin{align}
    \label{eqn:wave}
    &\Delta \mathbf{A}-\bm{\nabla}(\bm{\nabla}\mathbf{A})-\varepsilon\frac{\partial^2\mathbf{A}}{c^2\partial t^2}=-\frac{4\pi}{c}\mathbf{j},\\
    &\mathbf{j}=-ec\sum_j\bm{\alpha}_j\delta(\mathbf{r}_j-\mathbf{r}).
\end{align}

The wave equation and the interaction part of the Hamiltonian density can be expressed by means of the Fourier coefficients $\mathbf{A}(\mathbf{k})$ and $\mathbf{j}(\mathbf{k})$:

\begin{equation}
\begin{gathered}
    \label{eqn:wavek}
    \left(k^2-\mathbf{k}\otimes\mathbf{k}+\frac{\varepsilon}{c^2}\frac{\partial^2}{\partial t^2}\right)\mathbf{A}(\mathbf{k})=\frac{4\pi}{c}\mathbf{j}(\mathbf{k}),\\
    \mathbf{j}(\mathbf{k})=-\frac{ec}{(2\pi)^3}\sum_j\bm{\alpha}_j e^{-i\mathbf{k}\mathbf{r}_j},
\end{gathered}
\end{equation}

\noindent where $\mathbf{k}\otimes\mathbf{k}$ is the outer product operator.

\begin{equation}
    \label{eqn:intHamk}
    H_{int}=-(2\pi)^3 c^{-1} \int \mathbf{j}(-\mathbf{k})\mathbf{A}(\mathbf{k})d\mathbf{k}. 
\end{equation}

\section{\label{sec:2level}Two-level system approximation}
\subsection{\label{sec:sEleHam}Simplification of $H_e$}

Let us assume that radiation of the SASE process is in the resonance with a pair of channeled electron energy levels. This allows us to omit all the rest levels which are far from the resonance. As a result, the pair of the remaining states can be considered as  a two-level system. The first (ground state) is supposed to match the one possessing lower energy value whereas the second (excited state) matches the other one with higher energy. The complete structure of these states can be predicted using the group theory (see appendix \ref{appendix:structure} for details). In case of cubic crystal family and [001] axial channeling one of the states is twice-degenerate whereas the other is non-degenerate (it is not crucial which of the states is degenerate; in the present paper we assume that the second state is two-dimensional and has two orthogonal wave functions $\phi_{2,1}$ and $\phi_{2,2}$).  Introducing the vectors $\mathbf{d}_l=\int{\phi^*_1 \mathbf{r}_\bot \phi_{2,l} d\mathbf{r}_\bot}$, one can conclude that both $\mathbf{d}_1$ and $\mathbf{d}_2$ have the same absolute value $d$ and are perpendicular to each other.

The electron beam is supposed to be uniform. One can obtain the following expression for $H_e$:

\begin{multline}
    \label{eqn:eleHamDS}
    H_e(p_\parallel)=\gamma(p_\parallel)mc^2\ket{p_\parallel}\bra{p_\parallel}+\\+\frac{\hbar\Omega}{2}\bigg(\sum_{l=1,2}\ket{p_\parallel,2,l}\bra{p_\parallel,2,l}-\ket{p_\parallel,1}\bra{p_\parallel,1}\bigg),\\
    H_e=\int dp_\parallel H_e(p_\parallel),
\end{multline}

\noindent where $\hbar\Omega = \delta E_2-\delta E_1$. The first term in (\ref{eqn:eleHamDS}) describes the longitudinal motion of the electron whereas the other one is related to channeled energy levels.

\subsection{\label{sec:sj}Simplification of $\mathbf{j}(\mathbf{k})$}

Some simplifications can also be done in respect to the current density $\mathbf{j}(\mathbf{k})$ of channeled electrons. Considering the effect in the dipole approximation, one can deduce:

\begin{equation}
    \label{eqn:expsimp}
    \exp(-i\mathbf{k}\mathbf{r}_j)=\left(1-i\mathbf{k}_\bot\mathbf{r}_{\bot j}\right)e^{-i\left(k_\parallel r_{\parallel j}+\mathbf{k}_\bot \mathbf{r}_{0\bot j}\right)},
\end{equation}

\noindent where $r_{0\bot}$ is the position of a crystallographic axis; $e^{-ik_\parallel r_{\parallel}}$ results in the fact that only transitions with $p \rightarrow p-\hbar k_\parallel$ are allowed. Let us derive the matrix elements of $c\bm{\alpha}\exp \left( ik_\parallel r_{\parallel} \right)$ operator:

\begin{equation}
\begin{split}
    \label{eqn:alpha}
    &\bra{p,n}c\bm{\alpha}e^{ik_\parallel r_{\parallel}}\ket{q,n}=\mathbf{v}\delta(p-q+k_\parallel\hbar),\\
    &\bra{p,2,l}c\bm{\alpha}e^{ik_\parallel r_{\parallel}}\ket{q,1}=i\Omega\mathbf{d}_{l}\delta(p-q+k_\parallel\hbar),
\end{split}
\end{equation}

\noindent where $n$ describes a complete set of observables except for $p_\parallel$. Since we are interested only in the radiation resulting from channeling, one can omit the term with $\textbf{v}$. Let us find non-zero matrix elements of $\bm{\alpha}(\mathbf{k}_\bot\mathbf{r}_{\bot})\exp ik_\parallel r_{\parallel}$:

\begin{equation}
\begin{split}
    \label{eqn:alphakr}
    \bra{p,2,l}c\bm{\alpha}(\mathbf{k}_\bot\mathbf{r}_{\bot})e^{ik_\parallel r_{\parallel}}\ket{q,1}=\\=\mathbf{v}(\mathbf{k}\mathbf{d}_{l})\delta(p-q+k_\parallel\hbar).
\end{split}
\end{equation} 

In order to make further simplifications it is convenient to introduce the vector 

\begin{equation}
\begin{split}
    \label{eqn:vectora}
    \mathbf{a}_l(\mathbf{k})=(1+\frac{\mathbf{v}\otimes\mathbf{k}}{\Omega})\frac{\mathbf{b}_l}{b}.
\end{split}
\end{equation} 

\noindent and operator $\mathcal{S}(\mathbf{k})$ which performs transition $p \rightarrow p-\hbar k_\parallel$:

\begin{equation}
\begin{split}
    \label{eqn:Sdetermine}
    \mathcal{S}(\mathbf{k})=\int\ket{p-\hbar k_\parallel}\bra{p}dp.
\end{split}
\end{equation}

The Heisenberg's equation for (\ref{eqn:Sdetermine}) has the following form

\begin{equation}
\begin{split}
    \label{eqn:S}
    \frac{d\mathcal{S}(\mathbf{k})}{d t}=-\frac{i}{\hbar}[H,\mathcal{S}(\mathbf{k})]=i\mathbf{vk}\mathcal{S}(\mathbf{k}),
\end{split}
\end{equation}

\noindent with the solution

\begin{equation}
\begin{split}
    \label{eqn:Ssol}
    \mathcal{S}(\mathbf{k})=\mathcal{S}_0(\mathbf{k})\exp{ i\mathbf{vk}(t_0-t)},
\end{split}
\end{equation}

\noindent where  $\mathcal{S}_0$ is the value of operator $\mathcal{S}$ at $t=t_0$. Since we are interested in radiation caused by transitions between $\ket{n}$ and the influence of momentum change on channeling radiation is negligibly small, for the further investigation one can omit $\mathcal{S}_0$ and retain only the time-dependent part. Finally, the following expression for $\mathbf{j}(\mathbf{k})$ can be deduced:

\begin{equation}
\begin{split}
    \label{eqn:intHamS}
    \mathbf{j}(\mathbf{k})=-\frac{ie\Omega d}{(2\pi)^3}\sum_{j,l}(\mathbf{a}_l(-\mathbf{k})\ket{2,l}\bra{1}_j-\\
    -\mathbf{a}_l(\mathbf{k})\ket{1}\bra{2,l}_j) e^{-i\mathbf{k} (\mathbf{r}_{0j}+\mathbf{v}t)},
\end{split} 
\end{equation}

\noindent $\mathbf{r}_0=\bm{r}_{0\bot}-\bm{v}t_0$ can be interpreted as the initial position of an electron.

\subsection{Spin-like operators}

In order to simplify some notations let us introduce a new set of operators:

\begin{equation}
\begin{gathered}
    \label{eqn:spin}
    \sigma_{3l}=\frac{1}{2}(\ket{2,l}\bra{2,l}-\ket{1}\bra{1}),\\
    \sigma_{+l}=\ket{2,l}\bra{1},~\sigma_{-l}=\ket{1}\bra{2,l},\\
    \tau_{ll'}=\ket{2,l}\bra{2,l'}.
\end{gathered}
\end{equation}

The operators (\ref{eqn:spin}) satisfy the following commutation relations:

\begin{equation}
\begin{gathered}
    \label{eqn:spincom}
    [\sigma_{+l},\sigma_{-l}]=2\sigma_{3l},\\ [\sigma_{3l},\sigma_{-l}]=-\sigma_{-l},~[\sigma_{3l},\sigma_{+l}]=\sigma_{+l},\\
    [\sigma_{+1},\sigma_{-2}]=\tau_{12},~[\sigma_{-1},\sigma_{+2}]=-\tau_{21}.
\end{gathered}
\end{equation}

Substituting (\ref{eqn:spin}) in (\ref{eqn:intHamS}), one can deduce:

\begin{equation}
\begin{split}
    \label{eqn:spinJ}
    \mathbf{j}(\mathbf{k})=-\frac{ie\Omega d}{(2\pi)^3} \sum_{j,l}(\mathbf{a}_l(-\mathbf{k})\sigma_{+l,j}-\\-\mathbf{a}_l(\mathbf{k})\sigma_{-l,j}) e^{-i\mathbf{k} (\mathbf{r}_{0j}+\mathbf{v}t)}.
\end{split}
\end{equation}

The Heisenberg's equations for the new operators $\sigma_{-l}$ and $\sigma_{3l}$ have the form:

\begin{equation}
\begin{split}
    \label{eqn:sigma}
    &\frac{d\sigma_{-l}}{dt}=\frac{i}{\hbar}[H,\sigma_{-l}]=-i\Omega\sigma_{-l}-\frac{e\Omega d}{c\hbar}(2\sigma_{3l}+\\&+\tau_{ll'})\int\mathbf{a}_l(\mathbf{k})\mathbf{A}(\mathbf{k},\omega)e^{i\mathbf{k} (\mathbf{r}_{0j}+\textbf{v}t)-i\omega t}d\textbf{k}d\omega,\\
    &\frac{d\sigma_{3l}}{dt}=\frac{i}{\hbar}[H,\sigma_{3l}]=\frac{e\Omega d}{c\hbar}\int(\mathbf{a}_l(\mathbf{k})\sigma_{+l,j}+\\&+\mathbf{a}_l(-\mathbf{k})\sigma_{-l,j})\mathbf{A}(\mathbf{k},\omega)e^{i\mathbf{k} (\mathbf{r}_{0j}+\mathbf{v}t)-i\omega t}d\mathbf{k}d\omega.
\end{split}
\end{equation}

The Hermitian conjugation of the equation for $\sigma_{-l}$ leads to the equation for $\sigma_{+l}$.

\section{\label{sec:temp}Initial stage of SASE process}

Equations (\ref{eqn:sigma}) contain non-linear terms which make analytical calculations for operators hardly possible. However, at the initial stage of SASE process the electromagnetic field vector-potential and the change of occupations can be considered as small values and can be used as small parameters in order to apply the perturbation theory. 

\subsection{Zeroth-order approximation}

\begin{figure}[t]
\includegraphics[width=8cm]{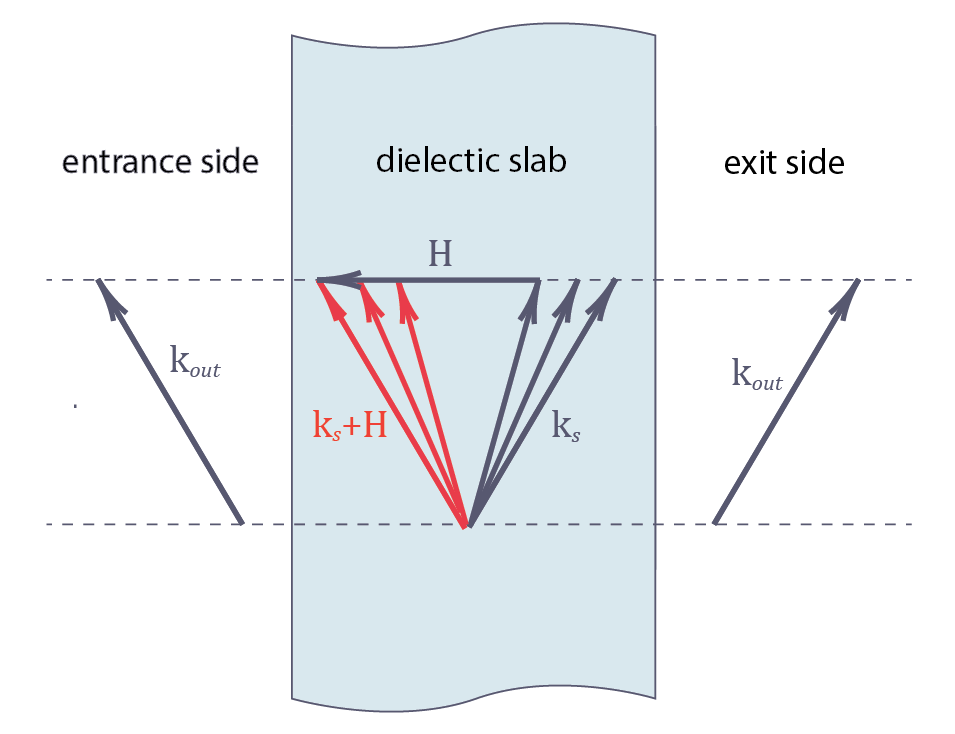}
\caption{Schematic distribution of wave vectors. Inside the dielectric slab: black vectors correspond to the solutions of (\ref{eqn:dispRel}), red vectors -- to the wave vectors related to the reflected waves. Outside the dielectric slab: black vectors correspond to $\mathbf{k}_{out}$. Dashed lines show the condition  $\mathbf{k}_{s\mathbf{r}}=\mathbf{k}_{out\mathbf{r}}$.}
\label{fig:wavevectors}
\end{figure}

In the zeroth approximation we assume:

\begin{align}
    \label{eqn:zerothsigma}
    \sigma_{-l}=\sigma_{-l0}\exp{i\Omega (t_0-t)},\\
    \mathbf{j}(\mathbf{k},\omega)=\frac{ie\Omega d}{(2\pi)^3} \sum_{j,l}(\mathbf{a}_l(\mathbf{k})\sigma_{-l0,j}\delta(\omega-\mathbf{k}\mathbf{v}-\Omega)-\nonumber\\
    -\mathbf{a}_l(-\mathbf{k})\sigma_{+l0,j}\delta(\omega-\mathbf{k}\mathbf{v}+\Omega)) e^{-i\mathbf{k}\mathbf{r}_{0j}}=\mathbf{j}_{sp}(\mathbf{k},\omega).
\end{align}

\noindent where $\sigma_{0}$ is $\sigma$ at $t=t_0$. In this case the resulting field is proportional to $\mathbf{j}_{sp}$. Assuming that particles propagate in perpendicular direction of the dielectric slab,  one can introduce the boundary condition $\mathbf{k}_{\bot}=\mathbf{k}_{\bot out}$, where $\mathbf{k}_{out}$ is the wave vector outside of the crystal (see figure \ref{fig:wavevectors}). Let $\alpha$ denote the angle between $\mathbf{k}_{out}$ and the electrons' direction. In the case of the first term of (\ref{eqn:zerothsigma}) one obtains

\begin{equation}
\begin{gathered}
    \label{eqn:zeroomega}
    \omega=\frac{\Omega}{1-\frac{v}{c}\cos{\alpha}},
\end{gathered}
\end{equation}

\noindent where we assume that $\varepsilon\approx1$.

Considering the peak value specified by the condition $\omega-\mathbf{k}\mathbf{v}=\Omega$, the second term in (\ref{eqn:zerothsigma}) can be omitted. The resulting current density $\mathbf{j}_{sp}$ describes the spontaneous radiation \cite{bird} and has zero quantum-mechanical average $\langle\sigma_{-0}\rangle$. However, $\langle\sigma_{-0}\sigma_{+0}\rangle$ isn't equal to zero and makes this current density to be similar to Langevin noise current.  

The peak photon density in the forward direction is

\begin{equation}
\begin{split}
    \label{eqn:pick in forward}
    \frac{dN}{d^2\mathbf{n}d\omega/\omega}=\frac{e^2\omega^2 L^2 P_e }{4\pi^2}\frac{\Omega^2d^2}{c^2},
\end{split}
\end{equation}

\noindent where $L$ is the thickness of the crystal slab, $P_e$ is occupation number of the excited channeling state.

\subsection{First-order perturbation theory}

In case of high current densities of the channeling electrons one has to consider the evolution of both radiation in a crystal and states of the channeled electrons self-consistently (it is important to stress that feedback of radiation towards electrons leads to the SASE process). 

Considering the first order perturbation theory one can find from (\ref{eqn:sigma}) that $\sigma_{-l}$ is proportional to the electromagnetic field vector-potential. As a result, one can introduce effective susceptibility due to linear response of the electrons' beam to the electromagnetic field, which has the following form in the Fourier space:

\begin{equation}
\begin{gathered}
    \label{eqn:sus}
    \chi(\mathbf{k},\omega)=\frac{\omega}{\omega-\mathbf{kv}-\Omega}\sum_l\chi_{b,l}\mathbf{a}_l(\mathbf{k})\otimes\mathbf{a}_l(\mathbf{k}),\\
    \chi_{b,l}=\frac{4\pi e^2 c n \Delta P_l}{\hbar\omega^3}\frac{\Omega^2d^2}{c^2},
\end{gathered}
\end{equation}

\noindent where $n$ is the density of electrons and $\Delta P_l=P_{e,l}-P_g$ is the difference between the population coefficients of excited and ground states. The possibility of population inversion and the value of $\Delta P_l$ can be found from the initial conditions. In order to simplify equations we consider that $\chi_l=\chi$, and $n$ can be expressed using beam brightness $B$ \cite{beams}. Accounting for the critical angle (\ref{eqn:critical}) one obtains:

\begin{equation}
\begin{gathered}
    \label{eqn:chicritical}
    n=\frac{B\pi\theta_c^2\gamma^2}{4ec},
    \chi_{b,l}=\frac{\pi^2 e B\theta_c^2\gamma^2 \Delta P_l}{\hbar\omega^3}\frac{\Omega^2d^2}{c^2},
\end{gathered}
\end{equation}

\noindent where $B$ is the brightness of the beam. Now let us take into account the crystallographic order. Its influence on the properties of the emitted field can be taken into account by means of periodic permittivity $\sum_j\chi_{H_j} e^{i\mathbf{H}_j\mathbf{r}}$. Here we consider that the emitted radiation is close to Bragg condition for reciprocal lattice vector $\mathbf{H}$, i.e. the two wave approximation.. It allows us to reduce permittivity and leave $\chi_{H} e^{i\mathbf{H}\mathbf{r}}+\chi_{-H} e^{-i\mathbf{H}\mathbf{r}}$ only. Finally, one can derive two equations for $\mathbf{k}$ and $\mathbf{k}_\mathbf{H}=\mathbf{k}+\mathbf{H}$:

\begin{figure*}[t]
\includegraphics[width=0.99\linewidth]{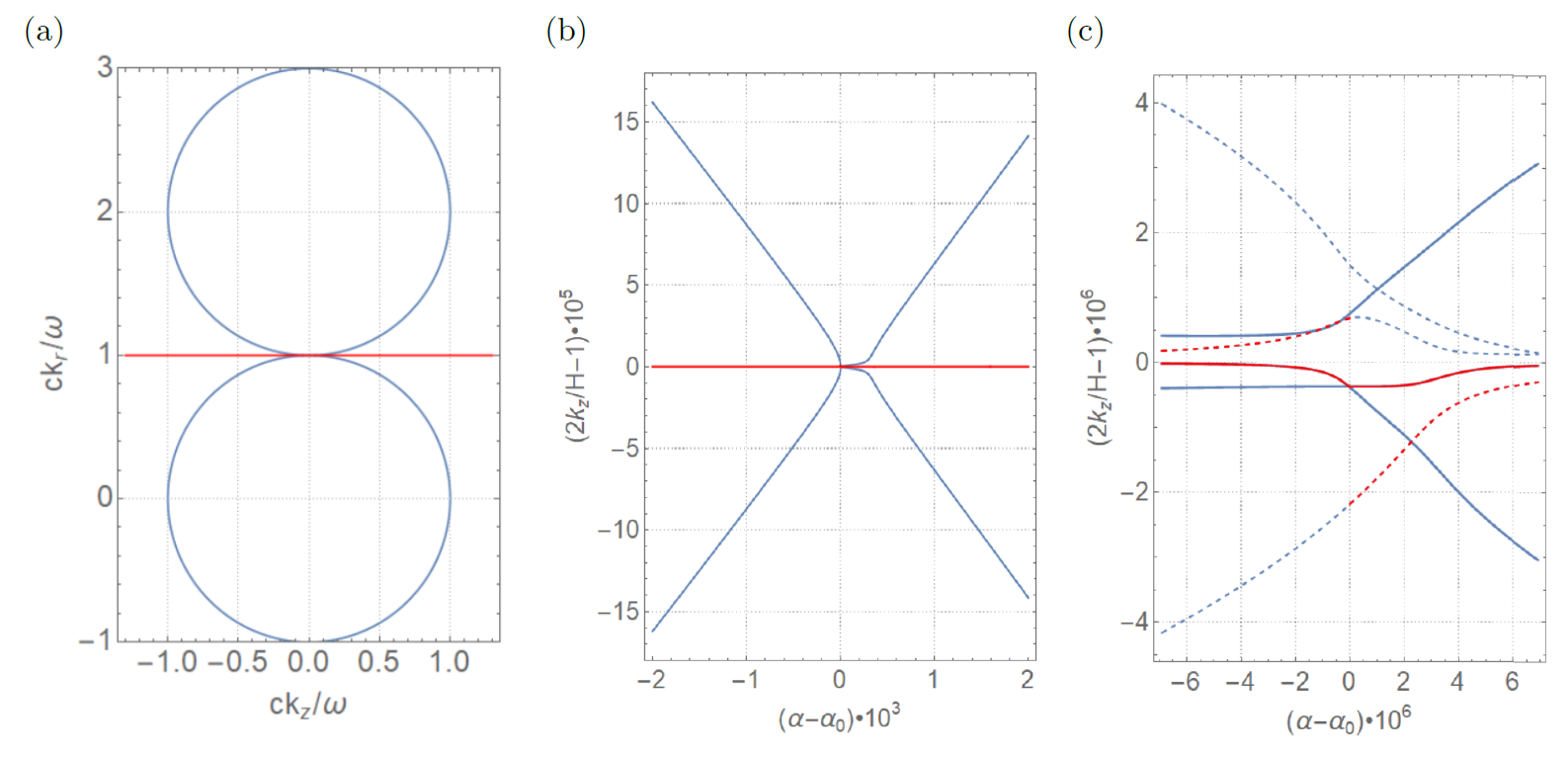}
\caption{Numerical results of dispersion surfaces shown in different scales corresponding to $\sigma$-polarization in case of Si crystal and resulting X-ray photon energy $\hbar \omega=4.5$ KeV, the electrons' energy  25 MeV, transition frequency $\Omega = 14.3$ eV, beam susceptibility $\chi_b=4.5\cdot10^{-13}$: (a) the roots of $X_\mathbf{k}X_{\mathbf{k+H}}-\frac{\omega^4}{c^4}\chi^2_H$ (blue lines) and $k_zv-\omega+\Omega$ (red line) in case of CR satisfying the Bragg condition; (b) the roots of (\ref{eqn:dispRel}) at $\omega_0$ in the vicinity of $\alpha_0$ where intersection of lines corresponding to different roots is shown (blue lines correspond to dynamic diffraction roots, red lines -- to CR condition); (c) a detailed view near the resonance condition (thick lines correspond to imaginary parts and dashed lines -- to real parts).}
    \label{fig:three}
\end{figure*}

\begin{equation}
\begin{split}
    \label{eqn:Aeqn}
    &\left(X_\mathbf{k}-\mathbf{k}\otimes\mathbf{k}-\frac{\omega^2}{c^2}\chi(\mathbf{k},\omega)\right)\mathbf{A}(\mathbf{k},\omega)-\frac{\omega^2\chi_{-H}}{c^2}\times\\&\times\mathbf{A}(\mathbf{k}_\mathbf{H},\omega)=\frac{4\pi}{c}\mathbf{j}_{sp}(\mathbf{k},\omega),
    \\&\left(X_{\mathbf{k}_\mathbf{H}}-\mathbf{k}_\mathbf{H}\otimes\mathbf{k}_\mathbf{H}\right)\mathbf{A}(\mathbf{k}_\mathbf{H},\omega)=\frac{\omega^2\chi_{H}}{c^2}\mathbf{A}(\mathbf{k},\omega),\\&X_\mathbf{k}=k^2-\frac{\omega^2}{c^2}\varepsilon_0(\omega),~\varepsilon_0(\omega)=1+\chi_0(\omega),
\end{split}
\end{equation}

\noindent where $\chi_0$ is the permittivity of the media. The vector $\mathbf{k}_\mathbf{H}$ is far from $\mathbf{k}$ which satisfies the condition $\omega-\mathbf{k}\mathbf{v}=\Omega$. Hence, in the second equation the terms containing $\mathbf{j}_{sp}(\mathbf{k},\omega)$ and $\chi(\mathbf{k}_\mathbf{H},\omega)$ are ommited.

\section{\label{sec:disp}Dispersion equation}

Let us derive the dispersion equation related to $\sigma$-polarization. First, let us introduce the unit vector $\mathbf{N}$ directed perpendicular to the surface of the crystal and the unit vector $\bm{\tau}$ directed along $\mathbf{N}\cross\mathbf{k}$. We also assume that $\mathbf{k}, \mathbf{N}$ and $\mathbf{v}$ lie in the same plane and in this case the dispersion equation can be factorized as a product of the dispersion equations corresponding to $\sigma$ and $\pi$ polarizations.  Multiplying equations (\ref{eqn:Aeqn}) by the vector $\bm{\tau}$,  one can derive the following dispersion equation for $\sigma$ polarization:

\begin{equation}
\begin{split}
    \label{eqn:dispRel}
    \left(X_\mathbf{k}X_{\mathbf{k+H}}-\omega^4 c^{-4} \chi_H\chi_{-H}\right)\left(\omega-\mathbf{kv}-\Omega\right)=\\=\chi_bX_{\mathbf{k+H}}\omega^3 c^{-2}.
\end{split}
\end{equation}

Derivation and analysis of the dispersion equation related to $\pi$ polarization is carried out in appendix \ref{appendix:pi}. Equation (\ref{eqn:dispRel}) can be utilized in order to determine the set of $\mathbf{k}$ that are close to relation $\omega-\mathbf{k}\mathbf{v}=\Omega$. Below we call them ``usual'' vectors. In order to get the wave vectors of the reflected waves (see figure \ref{fig:wavevectors}) one should add $\mathbf{H}$ to the roots of (\ref{eqn:dispRel}). The amplitudes of the reflected waves can be obtained by using the second equation from (\ref{eqn:Aeqn}). 
After finding the roots of the dispersion equation, one can write down the homogeneous solution of the wave field inside the crystal as:
\begin{figure*}[t!]
\includegraphics[width=0.99\linewidth]{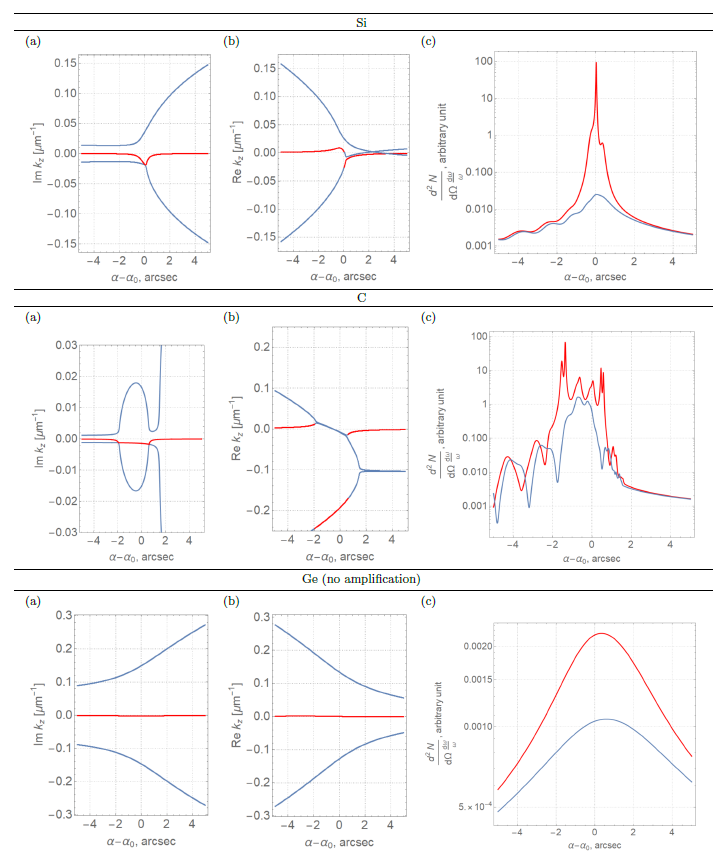}
\caption{Numerical simulation for $\omega=\omega_0$ case: (a) and (b) contain imaginary and real part of the roots of the dispersion equation for $\sigma$-polarization (blue lines correspond to dynamic diffraction roots, red lines -- to the CR condition); (c) the radiation intensity profile (red lines correspond to SASE, blue line -- to spontaneous radiation; the values are normalized over the peak value corresponding to the case $H = 0$ and $\chi_b$ = 0).}
\label{fig:three}
\end{figure*}
\begin{equation}
    \label{eqn:homo}
    \mathbf{A}(\mathbf{k},\omega)=\sum_sA_s(\mathbf{k},\omega)\mathbf{e}_s(\mathbf{k},\omega)\delta(k_z-k_{z,s}(\mathbf{k}_\mathbf{r},\omega)),
\end{equation}

\noindent where $s$ corresponds to the solutions of the dispersion equation, $A_s$ is a set of amplitudes that should be obtained from the boundary conditions, $\mathbf{e}_s$ is a set of polarization vectors, $\mathbf{k}_\mathbf{r}$ denotes the part of the vector $\mathbf{k}$ which is parallel to the surface of the crystal and $k_z$ denotes the remaining part, $k_{z,s}$ is a set of the dispersion equation roots.

It can be shown that the particular solution of (\ref{eqn:Aeqn}) equals zero. Since $\mathbf{j}_{sp}$ is proportional to $\delta(\omega-\mathbf{k}\mathbf{v}-\Omega)$, the particular solution is proportional to it as well. Multiplying the first equation by the expression $\omega-\mathbf{k}\mathbf{v}-\Omega$ and integrating it over $k_z$, one can derive that each term disappears except for the one with $\chi_b$. It leads to the particular solution being equal to zero which shows that the boundary conditions begin to play an important role in the SASE process.

\section{\label{sec:boundary}Boundary conditions}

Let us assume  that $\mathbf{H}$ is directed along $\mathbf{N}$ which means that $\mathbf{k}_{\bot}=\mathbf{k}_\mathbf{r}$ and $k_{\parallel}=k_z$. First of all, one should consider the boundary conditions for $\mathbf{A}(\mathbf{k},\omega)$ on both surfaces of the crystal. Taking into account that $\mathbf{A}(\mathbf{k},\omega)\cdot\mathbf{k}$ outside of the crystal is equal zero, one can derive the following condition for the entrance and exit surfaces:

\begin{equation}
\begin{gathered}
\label{eqn:Aboundary}
\mathbf{k}_{out}\left(1+\chi_0\mathbf{N}\otimes\mathbf{N}\right)\sum_s\mathbf{e}_sA_se^{ik_{z,s}l}=0,\\
\mathbf{k}_{out}\sum_s\mathbf{k}_s\cross\mathbf{e}_sA_se^{ik_{z,s}l}=0,
\end{gathered}
\end{equation}

\noindent where $l$ is the $z$-coordinate of the surface ($l=0$ for the entrance surface), $\mathbf{k}_{out}$ is a vector $\mathbf{k}$ which satisfies the relations  $\mathbf{k}_{s\mathbf{r}}=\mathbf{k}_{out\mathbf{r}}$. In case of the exit surface $k_{out~z}$ is positive and in the case of the entrance surface -- negative (see figure \ref{fig:wavevectors} for details).

For $\sigma$-polarization one can transform the second equation in (\ref{eqn:Aboundary}) as follows:

\begin{equation}
\begin{gathered}
\label{eqn:AboundarySimplified}
\sum_sA_s(k_{out~z}-k_{z,s})e^{ik_zl}=0,
\end{gathered}
\end{equation}

\noindent where we assume the absolute value of $\mathbf{e}_s$ to be a unity. 

In the case of the exit surface $|k_{out~z}-k_{z,s}| \ll H$ for roots of (\ref{eqn:dispRel}). However, for the reflected waves $k_{out~z}-k_{z,s} \approx H$, where $H$ is the modulus of $\textbf{H}$. Here we also assume that $\mathbf{k}_\mathbf{r}$ is close to $\mathbf{0}$. In case of the entrance surface one has the opposite situation: considering the second equation in (\ref{eqn:Aeqn}) the boundary conditions for $\sigma$ polarization turn to:

\begin{equation}
\begin{gathered}
\label{eqn:AboundarySimplified}
\sum_jA_j=0,~\sum_jA_j\frac{e^{ik_{z,j}L}}{X_{\mathbf{k}_j+\mathbf{H}}}=0,
\end{gathered}
\end{equation}

\noindent where $j$ counts usual waves only, $L$ is the thickness of the slab. The same results can be derived for $\pi$-polarization (see appendix \ref{appendix:pipolar} for details).

However, except for obtaining $A_s$ by using (\ref{eqn:AboundarySimplified}), one should also take into account the continuity equation for the current density on the first surface of the crystal:

\begin{equation}
\begin{split}
\label{eqn:jboundary}
\frac{ie\Omega d}{2\pi^2v}\sum_{j,l}\mathbf{a}_l(\mathbf{k}_{sp})\sigma_{-l0,j}e^{-i\mathbf{k}_{sp}\mathbf{r}_{0j}}=\\=-\frac{\omega^2}{c^2}\sum_jA_j\chi(\mathbf{k}_j,\omega)\mathbf{e}_j,
\end{split}
\end{equation}

\noindent where $\mathbf{k}_{sp}$ is a vector $\mathbf{k}$ satisfying the relations $\mathbf{k}_{s\mathbf{r}}=\mathbf{k}_{sp\mathbf{r}}$ and $\omega-\mathbf{k}_{sp}\mathbf{v}=\Omega$. One should note that equation (\ref{eqn:jboundary}) contains usual waves only.  

Considering the dispersion equation (\ref{eqn:dispRel}), one can derive the following the continuity equation for  the case of $\sigma$-polarization:

\begin{equation}
\begin{split}
\label{eqn:jboundarysigma}
\sum_j\frac{A_j}{X_{\mathbf{k}_j+\mathbf{H}}}\left(X_{\mathbf{k}_j}X_{\mathbf{k}_j+\mathbf{H}}-\frac{\omega^4}{c^4}\chi_H\chi_{-H}\right)=\\=-\frac{ie\Omega d}{2\pi^2v}\sum_{j}\sigma_{-0j}e^{-i\mathbf{k}_{sp}\mathbf{r}_{0j}}.
\end{split}
\end{equation}

The continuity equation in case of $\pi$-polarization is derived in appendix \ref{appendix:pipolar}. The set of equations composed of (\ref{eqn:jboundarysigma}) and (\ref{eqn:AboundarySimplified}) can be used to determine the wave amplitudes $A_j$. Since (\ref{eqn:jboundarysigma}) is the only equation which has nonzero r. h. s., $A_j$ turns out to be proportional to $\sigma_{-}$.   

Let us analyze the far field outside the crystal. Performing the Fourier transform of the Poynting vector and replacing the vectors with corresponding operators, one can derive the following equation for the photon density:

\begin{equation}
\begin{split}
    \label{eqn:phDens}
    \frac{dN}{d^2\mathbf{n}d\omega/\omega}=\bigg<\lim_{r\rightarrow\infty}\frac{r^2\omega^2}{2 c\hbar}\{\mathbf{A}^\dagger(r\mathbf{n},\omega),\mathbf{A}(r\mathbf{n},\omega)\}\bigg>,
\end{split}
\end{equation}

\noindent where $\{~...~,~...~\}$ denotes an anti-commutator and $\big<~...~\big>$ -- quantum-mechanical average. As it was shown above, $\mathbf{A}$ is proportional to $\sigma_{-}$ that has zero average value. However, expression (\ref{eqn:phDens}) contains the product of $\sigma_{-}$ and $\sigma_{+}$ which results in nonzero quantum-mechanical average. 

Using the Stratton~--~Chu equations, one can relate the field at the point of the far region to the field on the surface of the crystal in the following form:

\begin{equation}
\begin{split}
    \label{eqn:farToSurf}
    r\mathbf{A}(r\mathbf{n},\omega)=-\frac{2i\omega\pi}{c}\mathbf{nN}\left(1+\chi_0\mathbf{N}\otimes\mathbf{N}\right)\cdot\\ \cdot\sum_s\mathbf{e}_sA_se^{i(\frac{\omega}{c}r+k_{z,s}l)},
\end{split}
\end{equation}

\noindent where $\omega\mathbf{n}_\mathbf{r}=c\mathbf{k}_{s\mathbf{r}}$. The derivation of (\ref{eqn:farToSurf}) is carried out in appendix \ref{appendix:farToSurf}.

\section{\label{sec:optimization}Optimization}

Let us perform the optimization for the case of $\sigma$-polarization. Here we consider that $\chi_H=\chi_{-H}$.
In order to find out the conditions for when the amplification of radiation occurs, one should take into account that $e^{i\mathbf{kr}}$ leads to growth of the modulus when the imaginary part of $\mathbf{k}$ becomes negative. The goal of the present section is to formulate the conditions of maximal  value of the imaginary part of the root of the dispersion equation corresponding to CR. 

Let us consider equation (\ref{eqn:dispRel}) and include the Bragg condition into its  r. h. s.:

\begin{equation}
\begin{split}
    \label{eqn:dispRelRe}
    k_zv=\omega-\Omega-\frac{\chi_bX_{\mathbf{k+H}}\frac{\omega^3}{c^2}}{X_\mathbf{k}X_{\mathbf{k+H}}-\frac{\omega^4}{c^4}\chi^2_H}.
\end{split}
\end{equation}

The fraction in (\ref{eqn:dispRelRe}) is responsible for the increase of  the imaginary part  and can be considered as a perturbation to the root corresponding to CR at $\chi_b=0$. Therefore, one can substitute $k_z=(\omega-\Omega)/v$ in the r. h. s. Then the maximal value of the imaginary part can be obtained when the real part of the denominator is equal to zero.

We consider that the CR satisfies Bragg condition and, hence, the denominator of the fraction in (\ref{eqn:dispRelRe}) has two roots which are close to the CR root. Since the other roots are far from the CR case, for $\mathbf{k}_\mathbf{r}$ close to $\mathbf{0}$ one can reduce the values of $X_\mathbf{k}$ and $X_{\mathbf{k}+\mathbf{H}}$:

\begin{equation}
\begin{gathered}
    \label{eqn:reduced}
    X_\mathbf{k}\approx H(\delta k-\delta),~X_{\mathbf{k}+\mathbf{H}}\approx-H(\delta k+\delta),\\    \delta k=k_z-\frac{H}{2},~\delta=\sqrt{\frac{\omega^2}{c^2}(1+\chi_0)-\mathbf{k}^2_\mathbf{r}}-\frac{H}{2}.
\end{gathered}
\end{equation}

Let us consider the equation for the real part of the denominator (assuming that the imaginary part is much smaller than real one):

\begin{equation}
\begin{gathered}
    \label{eqn:realPart}
    H^2(\delta'^2-\delta k^2)=\frac{\omega^4}{c^4}\chi'^2_H,
\end{gathered}
\end{equation}

\noindent where $\delta'$ and $\delta''$ denote the real and imaginary parts of $\delta$, respectively. Since we have two parameters ($\omega$ and the direction of $\mathbf{k}$), one can introduce another condition maximizing the result. Accounting for (\ref{eqn:realPart}), the imaginary part of $k_z$ results in

\begin{equation}
\begin{split}
    \label{eqn:imPart}
    k''_z=-\frac{\chi_b\omega^3}{2vc^2H}\frac{\delta k+\delta'}{\delta'\delta''-\frac{\omega^4}{H^2c^4}\chi'_H\chi''_{H}}.
\end{split}
\end{equation}

From (\ref{eqn:reduced}) one can deduce that $\delta''\approx\frac{\omega\chi''_0}{2c}$. Setting the derivative in (\ref{eqn:imPart}) equal to zero, the second condition results in $\delta k=0$ when the roots of l.h.s. of (\ref{eqn:dispRel}) are mostly close to each other. Accompanied by the second condition, the expression (\ref{eqn:imPart}) turns into:

\begin{equation}
\begin{split}
    \label{eqn:finalIm}
     k''_z=-\frac{\chi_b\omega}{2c}\frac{1}{\chi_0''-\chi''_{H}},
\end{split}
\end{equation}

\begin{figure*}[t]
\includegraphics[width=0.99\linewidth]{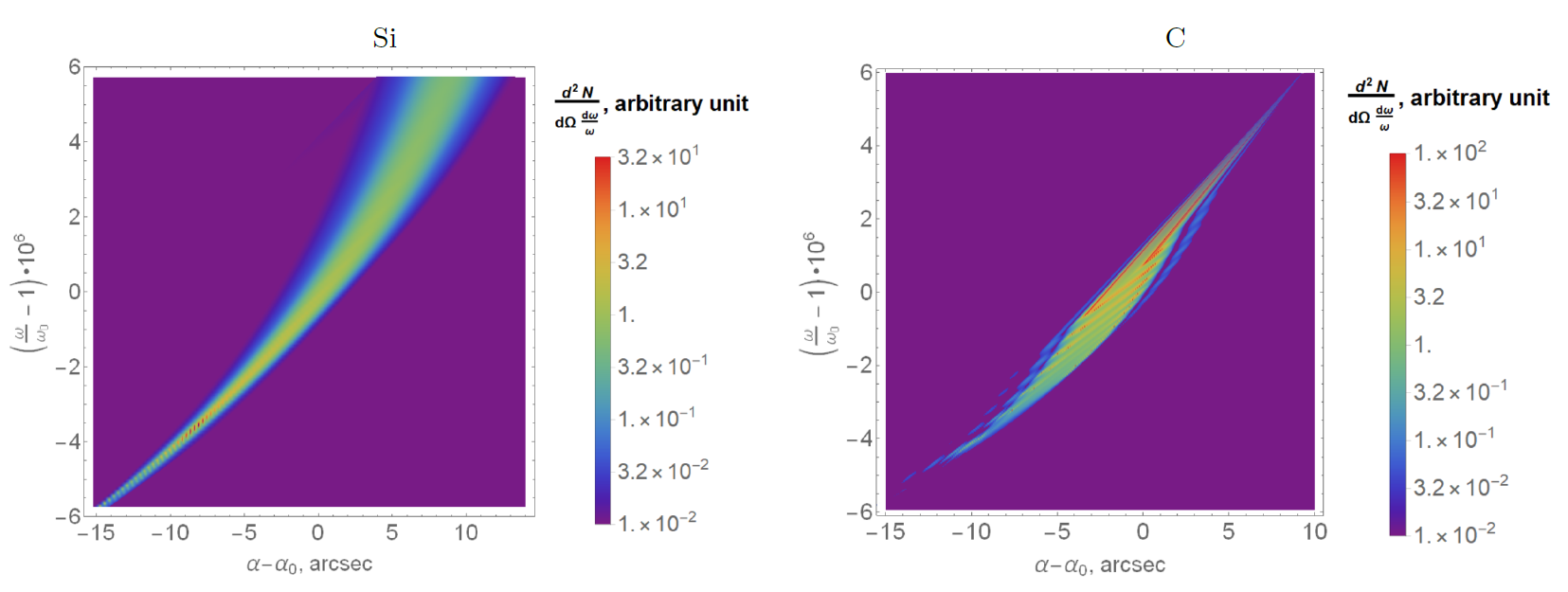}
\caption{Radiation intensity as a function of angle of  observation $\alpha$ and photon energy $\omega$ for Si and C crystals. The values are normalized over the peak photon density in forward direction corresponding to $H = 0$ and $\chi_b = 0$ case. } 
\label{fig:diagrams}
\end{figure*}

\begin{table*}[tb]
\begin{centering}
\caption{Parameters for different crystals. 
We consider that the electron beam brightness equals 
$ B = 1.7 \cdot 10^{19} \mathrm{A} \cdot \mathrm{m}^{-2}$ }
\begin{tabular}{c|cccccc}
\label{table:charTab}
Crystal & $\omega_0$, KeV & $\chi_h$                                  & $\chi_0$                                   & $\Delta P$ &%
$\chi_b$            & $\theta_c$, degree \\ 
\hline
Si      & 4.56      & $0.256\cdot10^{-4} + i 0.231\cdot10^{-5}$       & $-0.474\cdot10^{-4} + i 0.313\cdot10^{-5}$ & 3.3\%      &%
$3.6\cdot 10^{-13}$ &  0.184             \\
C       &6.94       & $0.111\cdot10^{-4} + i 0.640\cdot10^{-7}$       & $-0.304\cdot10^{-4} + i 0.915\cdot10^{-7}$ & 10.5\%     &%
$3.1\cdot 10^{-13}$ &  0.209 \\
Ge      & 4.38      & $0.592\cdot10^{-4} + i 0.597\cdot10^{-5}$       & $-0.993\cdot10^{-4} + i 0.857\cdot10^{-5}$ & 1.6\%      &%
$5.4\cdot 10^{-13}$ & 0.236  
\end{tabular}
\end{centering}
\end{table*}

\noindent where we assume that $H\approx2\omega/c$. One can see that taking into account Bragg diffraction leads to the increase of $|\Im{k_z}|$ (this effect is similar to anomalous transmission in Borrmann effect). 

Let us consider the values of the variables which leading to zero result in the resonance case. Using the condition $\delta k=0$, one can derive the expression for $\omega_0$:

\begin{equation}
    \label{eqn:omegaOpt}
     \omega_0=\frac{Hv}{2}+\Omega\approx\frac{H}{2}+\Omega-\frac{H}{4\gamma^2}.
\end{equation}

Let $\alpha$ be the angle between $\mathbf{v}$ and the wave vector outside the crystal $\mathbf{k}_{out}$ connected with the vector $\mathbf{k}$ inside. Using the relation (\ref{eqn:realPart}), one can derive the expression for $\alpha_0$:

\begin{align}
    \label{eqn:alphaOpt}
     \sin^2{\alpha_0}=\chi'_0-\chi'_H+1-\left(\frac{cH}{2\omega}\right)^2-\left(\frac{\omega\chi'_H}{cH}\right)^2\approx \nonumber \\
     \approx \chi'_0-\chi'_H+2\Omega-\frac{1}{\gamma^2}.
\end{align}

Figure \ref{fig:three} shows the solution of (\ref{eqn:dispRel}). One can see that there exists amplification of the generated radiation in case of Si and C crystals.

\section{\label{sec:num}Numerical results and discussion}

In this section we present numerical calculations performed for different crystals: Si, Ge and C. We consider axial channeling along [001] direction and the CR satisfying Bragg condition for 004 reflection at the Bragg angle close to $\pi/2$. For 25 MeV electrons moving along the channeling axis the occupation of the energy levels calculated within the Moliere potential results in the population inversion between the fourth level and degenerate second and third levels. The resulting X-ray photon energy, crystal susceptibilities, population inversion, beam susceptibility, Lindhard critical angle $\theta_c$ in each case are listed in the table \ref{table:charTab}. The  crystal slab  is supposed to be 200 $\mu$m thick. Assuming that the brightness of the electron beam is $B=1.7\cdot10^{19}$ A$\cdot\textmd{m}^{-2}$ \cite{beams}, the numerical results shown on figure \ref{fig:three} can be obtained. One can conclude that the resulting radiation in case of Si and C crystals experience  amplification. Despite of the fact that the beam susceptibility $\chi_b$ in the case of Ge is relatively big, the denominator of (\ref{eqn:finalIm}) is big as well leading to the absence of amplification in comparison to (\ref{eqn:pick in forward}).

Figure \ref{fig:diagrams} contains diagrams that illustrate the dependence of the radiation intensity on both angle $\alpha$ and resulting photon energy $\omega$ in the case of Si and C crystals. In case of C crystal we obtain smaller imaginary part of the crystallographic susceptibilities in comparison to the Si crystal that results in greater amplification. Hence, one can conclude that SASE in CR case  can lead to the amplification up to hundred times in the considered media. 

\begin{figure}[tb]
\includegraphics[width=0.99\linewidth]{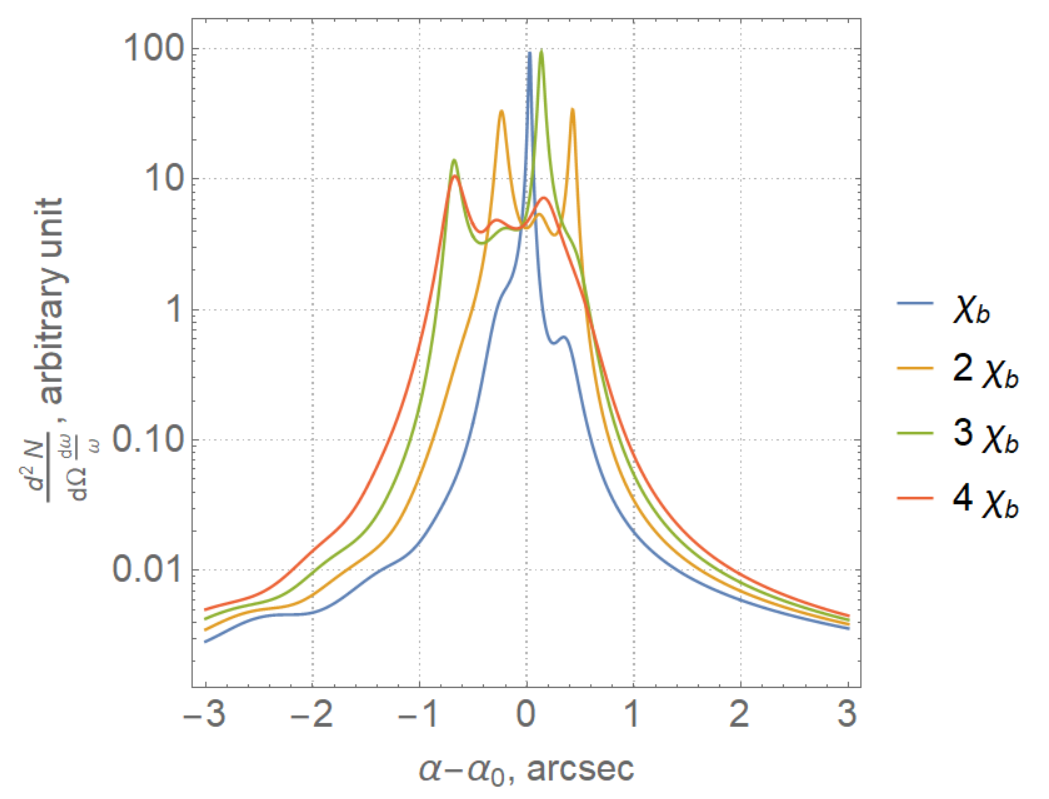}
\caption{Numerical simulation of the intensity profiles for various beam susceptibilities and Si crystal in case of 
$ \chi_b = 3.6 \cdot 10^{-13}$ 
and 
$ B = 1.7 \cdot 10^{19} \mathrm{A}\cdot \mathrm{m}^{-2}$. 
The values are normalized over the peak photon density in forward direction corresponding to the case  $H = 0$ and $\chi_b = 0$. }
\label{fig:on current}
\end{figure}

Finally let us consider the dependence of the results on the input current density. Figure \ref{fig:on current} shows the resulting intensity profile as a function of the electrons beam current density. One can see that the beam susceptibility change (that depends linearly on the current or brightness (\ref{eqn:chicritical})) leads to the broadening of the intensity profile.

\section{Conclusion}
\label{sec:conc}

The approach considered in the present paper allows one to give both qualitative and quantitative description of the initial stage of generation of SASE CR. This approach takes into account the influence of the periodical structure of the medium and even shows how to benefit from it. In order to describe this process beyond the exponential growth regime and more realistically one has to analyze the complete system (\ref{eqn:sigma}) and (\ref{eqn:wavek}). Another very important step is to take the dechanneling effect into account. These aspects can make a considerable contribution to the effect and will be the subject of our further investigations. It is important to note that SASE process requires high current densities and for a long time it was supposed to be impossible to implement this, especially in the case of crystals. But nowadays the required electron properties are feasible and the considered effect seems to be promising in terms of production the compact sources of bright and short X-ray pulses.

\section{Acknowledgements}
This work was supported by Belarusian Republican Foundation for Fundamental Research (BRFFR), grant No. F17RM-026 and Russian Foundation for Basic Research (RFBR), grant No. 17-52-04107.

\appendix

\section{\label{appendix:pi}Derivation of dispersion equations}

As it was mentioned above, vector $\bm{\tau}$ is perpendicular to $\mathbf{v}, \mathbf{k}$ and $\mathbf{H}$. Multiplying the expression (\ref{eqn:vectora}) by $\bm{\tau}$ one can deduce

\begin{equation}
    \mathbf{a}_l(\mathbf{k})\bm{\tau}=\frac{\bm{\tau}\mathbf{b}_l}{b},
\end{equation}

 \noindent so that due to degeneracy we obtain

\begin{equation}
\begin{split}
    \label{eqn:sigmaexp1}
    \bm{\tau}\chi(\mathbf{k},\omega)=\frac{\omega\chi_{b}\bm{\tau}}{\omega-\mathbf{kv}-\Omega}.
\end{split}
\end{equation}

Taking into account the properties of vector $\bm{\tau}$ and  multiplying each equation of (\ref{eqn:Aeqn}) by this vector and omitting the term with $\mathbf{j}_{sp}$, the following expressions can be deduced:

\begin{equation}
\begin{split}
    \label{eqn:sigmaexp2}
    &\left(X_\mathbf{k}-\frac{\omega^2}{c^2}\frac{\omega\chi_{b}}{\omega-\mathbf{kv}-\Omega}\right)\bm{\tau}\mathbf{A}(\mathbf{k},\omega)-\frac{\omega^2\chi_{-H}}{c^2}\times\\&\times\bm{\tau}\mathbf{A}(\mathbf{k}+\mathbf{H},\omega)=0,
    \\&X_{\mathbf{k}+\mathbf{H}}\bm{\tau}\mathbf{A}(\mathbf{k}+\mathbf{H},\omega)=\frac{\omega^2\chi_{H}}{c^2}\bm{\tau}\mathbf{A}(\mathbf{k},\omega).
\end{split}
\end{equation}

The solution of \eqref{eqn:sigmaexp2} leads to the dispersion equation for the case of $\sigma$-polarization:

\begin{equation}
\begin{split}
    \label{eqn:dispRelapp}
    \left(X_\mathbf{k}X_{\mathbf{k+H}}-\frac{\omega^4}{c^4}\chi_H\chi_{-H}\right)\left(\omega-\mathbf{kv}-\Omega\right)=\\=\chi_bX_{\mathbf{k+H}}\frac{\omega^3}{c^2}.
\end{split}
\end{equation}

As a result,  one can conclude that the dispersion equation can be factorized and $\sigma$-polarization doesn't affect $\pi$-polarization. This fact allows us to consider  the incident plane only. 
In this case $\chi(\mathbf{k},\omega)\mathbf{A}(\mathbf{k},\omega)$ can be reduced to:

\begin{equation}
\begin{gathered}
    \label{eqn:piexp1}
    \sum_l\mathbf{a}_l(\mathbf{k})\left(\mathbf{a}_l(\mathbf{k})\mathbf{A}(\mathbf{k},\omega)\right)=\mathbf{a}(\mathbf{k})\left(\mathbf{a}(\mathbf{k})\mathbf{A}_\pi(\mathbf{k},\omega)\right),\\
    \chi(\mathbf{k},\omega)=\frac{\omega\chi_b\mathbf{a}(\mathbf{k})\otimes\mathbf{a}(\mathbf{k})}{\omega-\mathbf{kv}-\Omega}, \mathbf{a}(\mathbf{k})=\mathbf{f}+\frac{\mathbf{v}k_\bot}{\Omega},
\end{gathered}
\end{equation}

\noindent where $\mathbf{f}$ is a unit vector directed perpendicular to $\mathbf{v}$ and laying in the plane with $\mathbf{k}$. Let us express $\mathbf{A}(\mathbf{k}_\mathbf{H},\omega)$ considering the second equation in (\ref{eqn:Aeqn})

\begin{equation}
\begin{split}
    \label{eqn:piexp2pr}
    \mathbf{A}(\mathbf{k}_\mathbf{H},\omega)=\frac{\chi_{H}}{\varepsilon_0}\bigg(X^{-1}_{\mathbf{k}_\mathbf{H}}(\mathbf{k}_\mathbf{H}^2-\mathbf{k}_\mathbf{H}\otimes\mathbf{k}_\mathbf{H})-\\-1\bigg)\mathbf{A}(\mathbf{k},\omega),
\end{split}
\end{equation}

\noindent and substituting this result into the first equation:

\begin{equation}
\begin{split}
    \label{eqn:piexp2}
     &\bigg(\frac{c^2X'_\mathbf{k}}{\omega^2\varepsilon_0}-\mathbf{k}\otimes\mathbf{k}-\frac{\omega^2}{c^2}\chi(\mathbf{k},\omega)-\frac{\omega^2\chi_{-H}\chi_{H}}{c^2\varepsilon_0X_{\mathbf{k}_\mathbf{H}}}\times\\&\times (\mathbf{k}_\mathbf{H}^2-\mathbf{k}_\mathbf{H}\otimes\mathbf{k}_\mathbf{H})\bigg)\mathbf{A}(\mathbf{k},\omega)=0,\\
     &X'_\mathbf{k}=\frac{\omega^2\varepsilon_0}{c^2}X_\mathbf{k}+\frac{\omega^4}{c^4}\chi_H\chi_{-H},
\end{split}
\end{equation}

\noindent where we omitted the current density term (as in the case of $\sigma$-polarization). Multiplying (\ref{eqn:piexp2}) by $\mathbf{k}_\mathbf{H}$ and $\mathbf{k}$, the following set of equations can be obtained:

\begin{equation}
\begin{split}
    \label{eqn:piexp2kkh}
     \bigg(\frac{c^2X'_\mathbf{k}}{\omega^2\varepsilon_0}\mathbf{k}_\mathbf{H}-(\mathbf{k}_\mathbf{H}\mathbf{k})\mathbf{k}-\frac{\omega^2}{c^2}\mathbf{k}_\mathbf{H}\chi(\mathbf{k},\omega)\bigg)\times\\\times\mathbf{A}(\mathbf{k},\omega)=0,\\
     \bigg(\frac{c^2X'_{\mathbf{k}_\mathbf{H}}}{\omega^2\varepsilon_0}\mathbf{k}-\frac{\chi_{-H}\chi_{H}}{\varepsilon_0^2}((\mathbf{k}_\mathbf{H}\mathbf{k})\mathbf{k}_\mathbf{H})+\\+\frac{X_{\mathbf{k}_\mathbf{H}}}{\varepsilon_0}\mathbf{k}\chi(\mathbf{k},\omega)\bigg)\mathbf{A}(\mathbf{k},\omega)=0,
\end{split}
\end{equation}

Calculating the determinant of the matrix corresponding to the set (\ref{eqn:piexp2kkh}) of equations, the dispersion equation for $\pi$-polarization can be derived:

\begin{equation}
\begin{split}
    \label{eqn:piexp3}
    &\mathcal{D}_\pi\left(\omega-\mathbf{kv}-\Omega\right)=\chi_b\frac{\omega^3}{c^2}\mathcal{F},\\
    &\mathcal{F}=\frac{\omega^2}{c^2}\bigg[\varepsilon_0 X_{\mathbf{k}_\mathbf{H}}(\mathbf{k}\times\mathbf{a}(\mathbf{k}))^2-\varepsilon_0\mathcal{D}_\sigma\mathbf{a}^2(\mathbf{k})-\\&-\frac{\omega^2}{c^2}\chi_H\chi_{-H}(\mathbf{k}_\mathbf{H}\times\mathbf{a}(\mathbf{k}))^2\bigg],\\
    &\mathcal{D}_\pi=X'_\mathbf{k}X'_{\mathbf{k}_\mathbf{H}}-\frac{\omega^4}{c^4}\chi_H\chi_{-H}(\mathbf{k}_\mathbf{H}\mathbf{k})^2,\\
    &\mathcal{D}_\sigma=X_\mathbf{k}X_{\mathbf{k}_\mathbf{H}}-\frac{\omega^4}{c^4}\chi_H\chi_{-H},
\end{split}
\end{equation}

where $\varepsilon_0=1+\chi_0(\mathbf{k},\omega)$.

\section{\label{appendix:pipolar}Boundary conditions in case of $\pi$ polarization}

Considering the first equation in (\ref{eqn:piexp2kkh}), one can deduce that the factor (written in brackets) is perpendicular to $\mathbf{A}(\mathbf{k},\omega)$. Moreover, taking into account that the second term in this factor has much greater value in respect to the other terms and is directed along $\mathbf{k}$, one can conclude that $\mathbf{A}(\mathbf{k},\omega)$ is perpendicular to $\mathbf{k}$ and $\mathbf{e}=\bm{\tau}\times\frac{\textbf{k}}{k}$  in case of $\pi$-polarization. Taking into account that $\mathbf{k}_\bot$ is close to $\mathbf{0}$, one can neglect the term containing $\chi_0$ in the first expression of set (\ref{eqn:Aboundary}). Finally, the following boundary condition can be deduced:

\begin{equation}
\begin{gathered}
\label{eqn:pipolarexp1}
\sum_sA_s(k_{out~z}-k_{z,s})e^{ik_zl}=0.
\end{gathered}
\end{equation}

One should stress that boundary conditions for the case of $\sigma$-polarization coincide with (\ref{eqn:pipolarexp1}).

Let us derive the continuity equation for the case of $\pi$-polarization. Taking into account the dependence of $\mathbf{a}(\mathbf{k})$  on $k_\bot$ only, (\ref{eqn:jboundary}) can be deduced to

\begin{equation}
\begin{split}
\label{eqn:pipolarexp2}
\frac{ie\Omega d}{2\pi^2v}\sum_{j,l}\sigma_{-l0,j}e^{-i\mathbf{k}_{sp}\mathbf{r}_{0j}}=\\=-\frac{\omega^3}{c^2}\sum_jA_j\frac{\chi_b\mathbf{a}(\mathbf{k}_j)\mathbf{e}_j}{\omega-\mathbf{k}_j\mathbf{v}-\Omega}.
\end{split}
\end{equation}

Finally, applying the dispersion equation (\ref{eqn:piexp3}) to (\ref{eqn:pipolarexp2}), we obtain

\begin{equation}
\begin{split}
\label{eqn:pipolarexp3}
\frac{ie\Omega d}{2\pi^2v}\sum_{j,l}\sigma_{-l0,j}e^{-i\mathbf{k}_{sp}\mathbf{r}_{0j}}=-\sum_jA_j\frac{\mathcal{D}_\pi}{\mathcal{F}}\mathbf{a}(\mathbf{k}_j)\mathbf{e}_j.
\end{split}
\end{equation}

\begin{table}[tb]
\caption{Character table of $C_{4v}$.}
\begin{tabular}{c|ccccc|c}
\label{table:charTab}
     $C_{4v}$ &E &$2C_4$ &$C_2$ &2$\sigma_v$ &2$\sigma_d$ &  \\
     \hline
     $A_1$ &1 &1 &1 &1 &1 &  \\
     $A_2$ &1 &1 &1 &$-1$ &$-1$ &  \\
     $B_1$ &1 &$-1$ &1 &1 &$-1$ &  \\
     $B_2$ &1 &$-1$ &1 &$-1$ &1 &  \\
     $E$ &2 &0 &$-2$ &0 &0 &$x,y$  \\
     
\end{tabular}
\end{table}
\begin{table}[tb]
\caption{Direct products of irreducible representations of $C_{4v}$.}
\begin{tabular}{c|ccccc}
\label{table:prodTab}
     $A_1$ &$A_2$ &$B_1$ &$B_2$ &$E$  \\
     \hline
     $A_2$ &$A_1$ &$B_2$ &$B_1$ &$E$  \\
     $B_1$ &$B_2$ &$A_1$ &$A_2$ &$E$  \\
     $B_2$ &$B_1$ &$A_2$ &$A_1$ &$E$ \\
     $E$ &$E$ &$E$ &$E$ 
     &\begin{tabular}{c}
     $A_1+A_2+$\\
    $+B_1 + B_2$
     \end{tabular}
     
\end{tabular}{}
\end{table}

\section{\label{appendix:farToSurf}Far field}

In this appendix we establish relations between the field values at the point of the far region with the field values on the crystal surface. Using the Stratton -- Chu equations, one can write down the following expression for $\mathbf{A}$ at any point outside the crystal:

\begin{equation}
\begin{split}
    \label{eqn:ftsexp1}
    \mathbf{A}(r'\mathbf{n},\omega)=\frac{1}{4\pi}\int_S\big[\mathbf{N}\times\left(\bm{\nabla}\times\mathbf{A}\right)\psi+\left(\mathbf{N}\times\mathbf{A}\right)\times\\\times\bm{\nabla}\psi+\left(\mathbf{N}\cdot\mathbf{A}\right)\bm{\nabla}\psi\big]ds,
\end{split}
\end{equation}

\noindent where $\psi=\exp(i\omega r/c)/r$ and $r$ is the distance between the point on the surface and the point outside the crystal. Integration is performed over the exit surface of the crystal.

In the case of far region one can write:

\begin{equation}
\begin{split}
    \label{eqn:ftsexp2}
    \bm{\nabla}\psi\approx \frac{i\omega}{c}\psi\mathbf{n},~\frac{1}{r}\exp(i\omega r/c)\approx \frac{C}{r'}\exp(-\frac{i\omega\mathbf{n}}{c} \mathbf{r}_\mathbf{r}),
\end{split}
\end{equation}

\noindent where $C$ is a constant with a unit absolute value. Since the radiation intensity is proportional to $|C|$, one can omit $C$ value below. Using the approximations introduced in (\ref{eqn:ftsexp2}), the expression (\ref{eqn:ftsexp1}) can be transformed to

\begin{equation}
\begin{split}
    \label{eqn:ftsexp3}
    r'\mathbf{A}(r'\mathbf{n})=\frac{1}{4\pi}\int_S\bigg[
    \left(\bm{\nabla}-\frac{i\omega\mathbf{n}}{c}\right)(\mathbf{A}\cdot\mathbf{N})+\\+\mathbf{N}\left(\mathbf{A}\cdot\frac{i\omega\mathbf{n}}{c}\right)-\left(\mathbf{N}\left(\bm{\nabla}+\frac{i\omega\mathbf{n}}{c}\right)\right)\mathbf{A}
    \bigg]\times\\\times\exp(-\frac{i\omega\mathbf{n}}{c} \mathbf{r}_\mathbf{r})ds.
\end{split}
\end{equation}

Expression (\ref{eqn:ftsexp3}) can be considered as a Fourier transform. Performing it for the first term leads to 0. Taking into account the value $\mathbf{n}\cdot\mathbf{A}(\frac{i\omega\mathbf{n}}{c})$ leads to 0 as well. Integrating the remaining terms in (\ref{eqn:ftsexp3}) and taking into account the boundary conditions results in

\begin{equation}
\begin{split}
    \label{eqn:ftsexp4}
    r'\mathbf{A}(r'\mathbf{n},\omega)=-\frac{2i\omega\pi}{c}\mathbf{nN}\left(1+\chi_0\mathbf{N}\otimes\mathbf{N}\right)\cdot\\ \cdot\sum_s\mathbf{e}_sA_se^{i(\frac{\omega}{c}r+k_{z,s}l)}.
\end{split}
\end{equation}

\section{\label{appendix:structure}Structure of the energy states}

The structure of both ground and excited states of the channeled electrons can be predicted by means of the group theory. Considering the dipole approximation, the transition between the levels are allowed if only the corresponding matrix elements  $\mathbf{r}_\bot$  are non-zero. As it was mentioned above, we consider the cubic crystal family and axial channeling in the direction [001]. Hence,  $H_e$ is invariant under $C_{4v}$ group transformations. Table \ref{table:charTab} contains the characters of $C_{4v}$. Since the components of $\mathbf{r}_\bot$ form the basis of $E$, one can determine that dipole transitions occur only between $A_1,A_2,B_1,B_2 \leftrightarrow E$ according to Table \ref{table:prodTab} and, therefore, one of the states is described by the 2-dimensional representation~$E$ whereas the other one -- by a 1-dimensional representation. It is not crucial which state is degenerate, so in this paper the second state is assumed to be 2-dimensional without loss of generality.  Let us determine $\phi_{2,2}$ as a result of the action of $C_4$ on $\phi_{2,1}$ and $\mathbf{d}_l=\int{\phi^*_1 \mathbf{r}_\bot \phi_{2,l} d\mathbf{r}_\bot}$. Other matrix elements are equal to zero due to symmetry properties. In this case $\mathbf{d}_1$ and $\mathbf{d}_2$ have the same value $d$ but are perpendicular to each other.

%

\end{document}